\pdfoutput=1

\documentclass{article}
\usepackage{amsmath,amssymb,enumerate,eucal,pdfpages,theorem,xcolor}

\numberwithin{equation}{section}

\theoremheaderfont{\normalfont\scshape}
{\theorembodyfont{\rmfamily}\newtheorem{dfn}{Definition}[section]}
{\theorembodyfont{\rmfamily}\newtheorem{exm}{Example}[section]}
{\theorembodyfont{\rmfamily}\newtheorem{rmk}{Remark}[section]}

\newtheorem{thm}{Theorem}[section]

\def\mn{\medskip\noindent}

\def\sn{\smallskip\noindent}

\title{Amalgamated Geometric Structure of the Local Multiverse}
\author{Igor Yu. Potemine \\ 
\small Institut de Mathématiques, Universit\'e Paul Sabatier, Toulouse, France}
\date{}

\begin{document}
\maketitle
\abstract{We consider \emph{multiverses} as time-amalgamated multiply warped products of Lorentzian (Einstein) manifolds. We define the \emph{Local Multiverse} as timely-connected component of our physical $(3+1)$-spacetime. It is a collection of ``parallel universes" with (mutually) synchronized timelines. Metaphysical considerations suggest that the Local Multiverse could be an extremely complex agglomeration with, at least, several hundred parallel universes in the Solar neighbourhood (and many thousands in galaxy bulks). In this paper we study a simplified time-almagamated globally hyperbolic model. Our picture implies the multiversality of elementary particles which are, actually, transcosmic (super)strings with multiple endpoints on parallel universes considered as $D$-branes.}

\medskip

\tableofcontents

\section{Introduction}

By definition, a \emph{spacetime} or \emph{universe} $(X,g)$ in this paper means a connected time-oriented $(n+1)$-dimensional Lorentzian manifold of signature $(1,n)$. The product of two universes with dimensions $m+1$ and $n+1$ gives a pseudo-Riemmanian manifold of signature $(2,m+n)$. 

\smallskip However, when timelines of two universes are synchronized, we can take time-amalgamated products and coproducts. It leads to the natural definition of \emph{multiverses} as such amalgamated (co)products. Basically, there is a unique timeline (up to appropriate synchronizations) in all parallel universes of the same multiverse.

\smallskip The \emph{Local Multiverse} is a (time-)connected component of our physical $(3+1)$-spacetime in the collection of all universes. According to Buddhist and Hindu cosmologies and other metaphysical considerations, we might suppose that the Local Multiverse is a huge agglomeration of parallel universes.

\section{Globally hyperbolic multiverses}

Let us denote $\mathbb{T}$ the 1-dimensional timeline $\mathbb{R}^1$, considered as a manifold with negative definite metric $-dt^2$. A \emph{globally hyperbolic universe} is a spacetime $(X,g)$ admitting a global time function $\tau:X\rightarrow \mathbb{T}$ (\emph{cf.}~\cite{Haw}) as well as a \emph{Cauchy hypersurface} $C=\tau^{-1}(\{0\})$ (\cite[\S 6.6]{HE}, \cite[\S 3.2]{BEE}). Recall that every inextendile non-spacelike curve intersects $C$ exactly once. Global time functions are continuous and strictly increasing along the future-directed non-spacelike curves. Normally, we will consider \emph{Cauchy time functions} giving foliations of $X$ by Cauchy hypersurfaces $\tau^{-1}(\{c\})$, $c\in \mathbb{T}$.

\begin{dfn}
Let $X_1,\dots,X_k$ be a collection of globally hyperbolic spacetimes of dimensions $n_i^{}+1$ with time functions $\tau_i^{}:X_i\rightarrow \mathbb{T}_i=\mathbb{T}$, $1\leqslant i\leqslant k$. Then the pullback / fibered product
\begin{equation}
\mathbf{\Pi}_{\mathbb{T}}^{} X_i = X_1\times_{\mathbb{T}}^{} X_2\times_{\mathbb{T}}^{}\cdots \times_{\mathbb{T}}^{} X_k
\end{equation}
in the category of topological manifolds will be called a \emph{globally hyperbolic multiverse} associated with $(X_1,\dots,X_k)$ and $(\tau_1^{},\dots,\tau_k^{})$. Pullbacks of type (2.1) will be also called
\emph{time-amalgamated products}.
\end{dfn}

\begin{rmk}The dual notion is the \emph{amalgamated sum} of universes along synchronized timelines.
\end{rmk}

By \emph{Geroch's splitting theorem} \cite{Ger} (improved by Bernal-S\'anchez \cite{BS1,BS2}) each $X_i$, $1\leqslant i\leqslant k$, is homeomorphic (diffeomorphic) to $\mathbb{T}\times C_i$ where $C_i$ is a (smooth) Cauchy hypersurface in $X_i$. Consequently, $M$ is homeomorphic (diffeomorphic) to $\mathbb{T} \times S_1\times\dots\times S_k$.

\begin{exm} 
\emph{(Naive Minkowski-de Sitter multiverse)} Consider the $(n+1)$-di\-mensional Minkowski spacetime $\mathbb{R}^{1,n}$ with metric
\begin{equation}
-dt^2+dx_1^2+\cdots+dx_n^2
\end{equation}
and de Sitter spacetime $\mathbb{T}\times S^3$ with warped metric 
\cite[\S 5.3]{BEE}
\begin{equation}
-dt^2+r^2\cosh^2(t/r)h
\end{equation}
where $h$ is the usual Riemannian metric on $S^3$ with constant sectional curvature $K=+1$. Then the associated multiverse is homeomorphic (diffeomorphic) to $\mathbb{T}\times\mathbb{R}^n\times S^3$ with metric
\begin{equation}
-dt^2+[dx_1^2+\cdots+dx_n^2]+[r^2\cosh^2(t/r)h]\,.
\end{equation}
\end{exm}

Strictly speaking, even in simple splitted cases, our globally hyperbolic multiverses should not be considered as universes. It's rather a collection of universes with their own internal physical theories, connected by an ambient interversal space filled by transcosmic strings 
(\emph{cf.}~sect.~7).

\section{Improved Geroch's splitting theorem}

In this section, we state an improved version of Geroch's splitting theorem, due to Bernal and S\'anchez \cite[theorem 1.1]{BS2}.

\begin{thm}
Let $(X,g)$ be a globally hyperbolic spacetime. Then, it is isometric to the smooth product manifold
\begin{equation}
\mathbb{T}\times S,\  g = -\beta(t,x)dt^2 + \widetilde{g}
\end{equation}
where $S$ is a smooth spacelike Cauchy hypersurface, the natural projection 
$\tau:\mathbb{T}\times S\rightarrow \mathbb{T}$ is a smooth time function with past-directed timelike gradient $\nabla\tau$, $\beta:\mathbb{T}\times S\rightarrow\, ]0;+\infty[$ is a smooth function and $\widetilde{g}$ is a Riemannian metric on each slice $S_c=\tau^{-1}(\{c\})$, $\forall c\in\mathbb{T}$.
\end{thm}

\begin{rmk} 1) \smallskip Bernal and S\'anchez propose to call \emph{temporal} those time functions that appear in this theorem.

\sn 2) The proof extends directly from dimension $3+1$ to any dimension $n+1$.
\end{rmk}

\section{Multiply warped Lorentzian metrics}

We have defined globally hyperbolic multiverses in the category of topological manifolds. However, the smooth version of the Geroch's splitting theorem allows us to define multiply warped Lorentzian metrics on multiverses. We adopt an appropriate version of the standard definition going back to Bishop-O'Neill (\emph{cf.}~\cite[\S 3.6]{BEE}).

Let $X=\mathbb{T}\times S_1$ and $Y=\mathbb{T}\times S_2$ be two (splitted) spacetimes with metrics $g=-\beta_1dt^2 +\widetilde{g}$ and $h=-\beta_2dt^2 +\widetilde{h}$ resp. We can endow the multiverse $X\times_{\mathbb{T}}^{} Y$ with the \emph{doubly warped Lorentzian metric}
\begin{equation}
g\times_{(\beta_1,\beta_2)}^{} h = g\,_{\beta_1}^{}\!\!\times
_{\beta_2}^{} h = -\beta_1\beta_2dt^2 +\beta_2\widetilde{g}+\beta_1\widetilde{h}
\end{equation}

\smallskip Inductively, we can also define \emph{multiply warped Lorentzian metrics} on any globally hyperbolic multiverse.

\smallskip Actually, the definition can be extended to more general splitted cases and partial Cauchy hypersurfacess. For instance, the universal anti-de Sitter space is strongly causal, but not globally hyperbolic. Indeed, it has no global Cauchy hypersurfaces 
\cite[\S 5.2]{HE}. Nonetheless, one can consider multiverses containing AdS spaces.

\begin{exm} 
\emph{(Minkowski-AdS$_4$ multiverse)} The metric of the anti-de Sitter spacetime AdS$_4$, in appropriate coordinates $(t',r,\theta,\phi)$ can 
be written in the form \cite[\S 5.3]{BEE}:
\begin{equation}
ds^2 = -\cosh^2(r)(dt')^2+dr^2+\sinh^2(r)(d\theta^2+\sin^2\theta
d\phi^2)
\end{equation}
Thus, the \emph{Minkowski-AdS$_4$ multiverse} $\mathbb{R}^{1,n}
\times_{\mathbb{T}}^{}\mathrm{AdS}_4$ has the warped metric
\begin{align}
-\cosh^2(r)(dt')^2 &+[\cosh^2(r)\left(dx_1^2+\cdots+dx_n^2\right)]+\\
&+[dr^2+\sinh^2(r)\left(d\theta^2+\sin^2\theta d\phi^2\right)].
\end{align}
\end{exm}

\smallskip Finally, notice that metric (4.1) is conformal to
\begin{equation}
ds^2=-dt^2 +\frac{\widetilde{g}}{\beta_1}+\frac{\widetilde{h}}{\beta_2}\,.
\end{equation}

\section{Multiversal FLRW models}

\smallskip Recall that \emph{Friedmann-Lemaître-Robertson-Walker (FLRW) universe} is a spacetime $X=\mathbb{T}\times S$, where $S$ has constant sectional curvature $K$, with metric of type:
\begin{equation}
ds^2=-dt^2+a(t)^2d\sigma^2\,.
\end{equation}
Here $a(t)$ should satisfy the so-called \emph{Friedmann equations}
\cite[\S 5.3, (5.12) and (5.13)]{HE}, giving a solution to \emph{Einstein equations}.
\begin{dfn}
Any time-amagamated products of FLRW universes will be called \emph{FLRW multiverses}.
\end{dfn}
The basic example is a time-amalgamated product of \emph{Einstein static universes} $\mathbb{T}\times S^n$, where $S^n$ is the $n$-sphere with standard spherical Riemannian metric \cite[Ex.~5.11]{BEE}. So, such \emph{Einstein static multiverses} looks like bouquets of spheres, remaining static when time evolves.

\smallskip Notice also that Minkowski, de Sitter and Anti-de Sitter spaces can be conformally embedded into the Einstein static universe. However, it requires time modifications $t\rightarrow t'$ and horizontal slices, corresponding to $\{t'=\mathrm{const}\}$, are not Cauchy hypersurfaces in general \cite[ch.~5, fig.~14-21]{HE}.

\section{Models with multiple Big Bangs and Big Crunches}

We can extend our initial definition 2.1 to the case where $\mathbb{T}_i$ are open intervals (finite or infinite) of the real timeline. So, the pairwise amalgamation of universes $X_i$ and $X_j$ happens only in the intersection $\mathbb{T}_i\cap\mathbb{T}_j$.

\smallskip Let $\Lambda_{\mathrm{crit}}$ be the critical value of the cosmological constant $\Lambda$. The nature of FLRW universes depends on 
$\Lambda$ (see \cite[\S 5.3]{HE}).

\includepdf[pages=1, scale=.4]{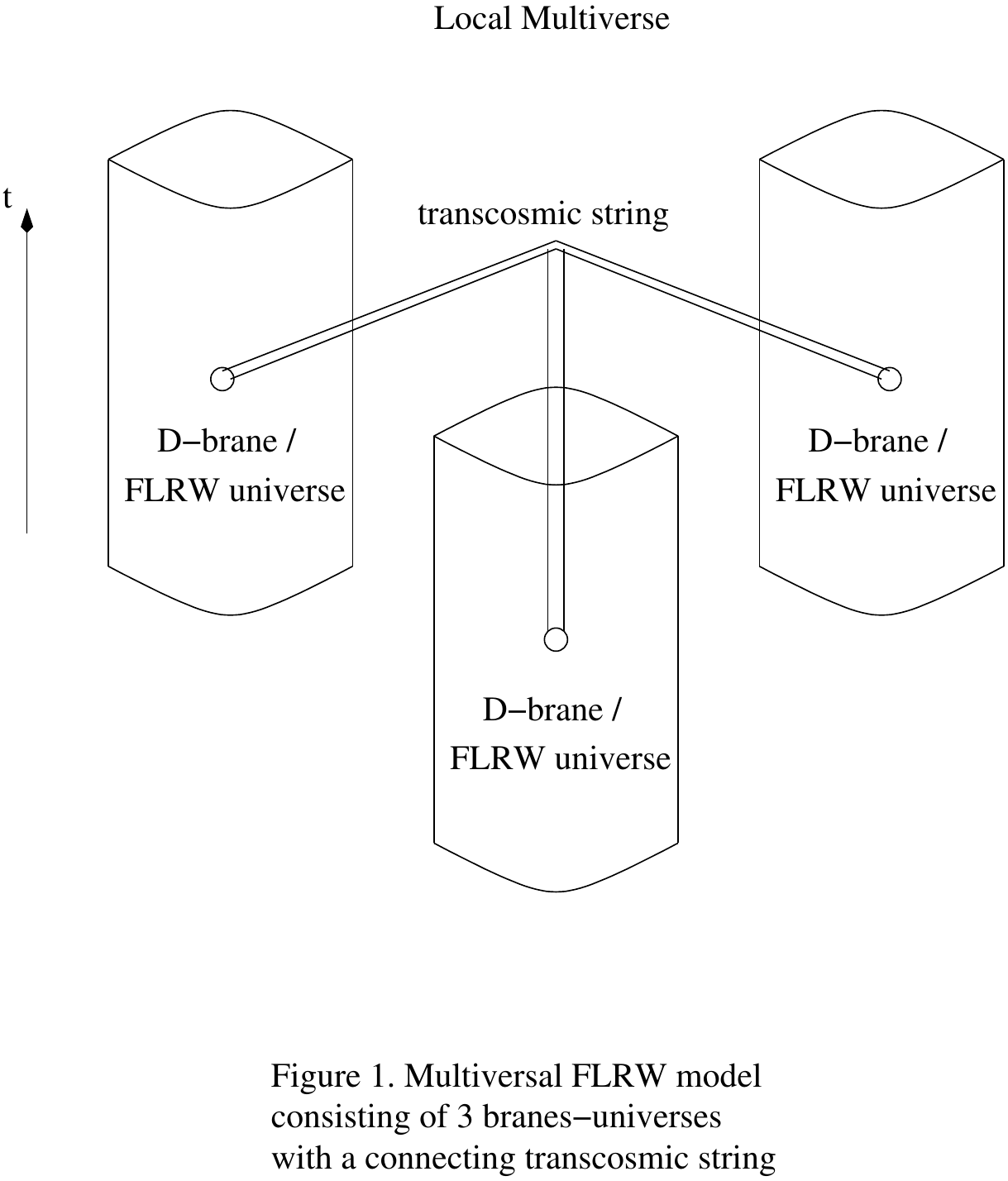}

\begin{itemize}
\item ($\Lambda<\Lambda_{\mathrm{crit}}$) There are FLRW universes expanding from an initial singularity (``Big Bang") and recollapsing to a second singularity (``Big Crunch").

\item ($\Lambda=\Lambda_{\mathrm{crit}}$) There are FLRW universes starting from an initial singularity (``Big Bang") and asymptotically approaching the Einstein static universe.

\item ($\Lambda>\Lambda_{\mathrm{crit}}$) There are FLRW universes expanding forever from an initial singularity (``Big Bang") and asymptotically approaching the so-called \emph{steady state model} 
\cite[\S 5.2]{HE}.
\end{itemize}
This list is not exhaustive and there are also different possibilities and, in particular, other options for Big Crunches.

\smallskip Amalgamating such FRLW universes (possibly with different cosmological constants), one obtains multiversal FLRW models with multiple singularities. These singularities correspond to attachments and detachments of universes to/from the Local Multiverse.

\section{Matrix of transcosmic superstrings}

We suppose that universes, belonging to the same multiverse 
(\emph{e.g.}~Local Multiverse), are interconnected by myriads of transcosmic (super)strings. At least, all baryons are supposed to be highly multiversal, i.e. existing simultaneously in numerous universes. In other words, observed baryonic particles are just endpoints of transcosmic baryonic strings.

\begin{rmk} Strictly speaking, such general theory of open strings \cite{Pol} with unrestricted numbers of endpoints on multiple D-branes \cite{Joh} does not yet exist. So, our considerations in this section are informal and hypothetical.
\end{rmk}

\smallskip In this perspective, the \emph{Local Multiverse} can be alternatively defined as the common world volume of all transcosmic strings with endpoint particles in our physical universe.

\smallskip Here we speak mainly about baryons of the first generation (protons and neutrons) as principal constituents of the matter. This is also related to the \emph{Matrix of the Local Multiverse}, but it is a different story.

\section{Conclusion}

In this paper we sketched an amalgamated geometric construction of multiverses, starting with globally hyperbolic models and giving simple examples. It permits to study FLRW multiverses with multiple Big Bangs and Big Crunches. However, these events are not catastrophic and correspond to attachments and detachments of universes from the Local Multiverse.

\smallskip In terms of a generalized superstring theory, the Local Multiverse is the common world volume of transcosmic strings having endpoints in our universe. These endpoints are supposed to be baryonic particles of the first generation (protons and neutrons) as principal constituents of the matter and of so-called Matrix of the Local Multiverse.

\smallskip Basically, this \emph{Multiversal Matrix} is a collection of transcosmic baryonic strings, sealed by electroweak and gluonic fields, but this topic is well beyond the scope of the present paper.

\smallskip Concerning the number of ``parallel universes" inside the Local Multiverse, I could refer only to my metaphysical book \cite{Pot}. It is supposed to be $250+$ on our (multiversal!) planet Earth, several hundreds in the Solar neighborhood and, at least, several thousands in star clusters and galaxy bulks. 

\smallskip Hopefully, experimental physicicts would be able to prove the multiversality of elementary particles during this century.

\mn {\scriptsize \textbf{Acknowledgments.} I would like to thank my colleague Joseph Tapia for fruitful and inspiring discussions.}

\bibliographystyle{unsrt} 
\bibliography{locmult1}

\bigskip
\begin{flushright}
Igor Potemine\\
Institut de Mathématiques\\
Universit\'e Paul Sabatier\\
118, route de Narbonne\\
31062 Toulouse (France)
\end{flushright}

\begin{flushright}
e-mail: igor.potemine@math.univ-toulouse.fr
\end{flushright}

\end{document}